\documentclass[final,5p,twocolumn,times,sort&compress]{elsarticle}

\usepackage{epsfig}

\def\al{\alpha}
\def\be{\beta}

\def\de{\delta}
\def\ep{\epsilon}

\def\th{\theta}

\def\la{\lambda}

\def\ta{\tau}

\def\ch{\chi}

\def\De{\Delta}

\def\half{{\textstyle{1\over 2}}}
\def\ol{\overline}

\def\lsim{\mathrel{\rlap{\lower4pt\hbox{\hskip1pt$\sim$}}
    \raise1pt\hbox{$<$}}}
\def\gsim{\mathrel{\rlap{\lower4pt\hbox{\hskip1pt$\sim$}}
    \raise1pt\hbox{$>$}}}

\def\etal{{\it et al.}}

\newcommand{\beq}{\begin{equation}}
\newcommand{\eeq}{\end{equation}}
\newcommand{\bea}{\begin{eqnarray}}
\newcommand{\eea}{\end{eqnarray}}
\newcommand{\rf}[1]{(\ref{#1})}

\def\ring#1{{\mathaccent'27 #1}}
\def\cri{\ring{c}}
\def\ari{\ring{a}}

\def\msm{3$\nu$SM} 

\def\h{h^\nu_{\rm eff}}
\def\hb{h^{\overline\nu}_{\rm eff}}
\def\ue{U(E)}
\def\ueb{\ol U(E)}
\def\le{L_{a'b'}(E)}
\def\leb{\ol L_{a'b'}(E)}

\def\lto{L_{31}}
\def\ltob{\ol L_{31}}
\def\ltt{L_{21}}
\def\lttb{\ol L_{21}}

\def\cc{\cri}
\def\aa{\ari}

\begin{document}

\begin{frontmatter}

\title{Three-parameter Lorentz-violating texture for neutrino mixing}

\author{Jorge S.\ D\'iaz and V.\ Alan Kosteleck\'y}

\address{Physics Department, Indiana University,
Bloomington, IN 47405, U.S.A.}

\address{}
\address{\rm IUHET 552, December 2010;
accepted for publication in Physics Letters B}

\begin{abstract}

A three-parameter model of neutrino oscillations 
based on a simple Lorentz- and CPT-violating texture
is presented. 
The model is consistent with established data 
and naturally generates 
low-energy and neutrino-antineutrino anomalies of the MiniBooNE type.
A one-parameter extension incorporates the MINOS anomaly,
while a simple texture enhancement accommodates 
the LSND signal.

\end{abstract}

\end{frontmatter}

In the minimal Standard Model (SM) of particle physics,
the three types of neutrinos are massless 
and preserve flavor as they propagate.
However,
compelling evidence now exists for neutrino flavor oscillations,
including the confirmed disappearance
of solar, reactor, atmospheric,
and accelerator neutrinos.
The canonical explanation for these oscillations
assumes that the three known flavors of neutrinos 
have a tiny mass matrix 
with nondiagonal components.
In the usual extension of the SM
to three flavors of massive neutrinos (\msm),
the $3\times3$ matrix governing oscillations 
involves two mass-squared differences,
three mixing angles,
and a CP-violating phase.
For suitable values of these six parameters,
the \msm\ successfully describes established oscillation data
\cite{pdg}.

In recent years,
several experiments have adduced some evidence 
for anomalous neutrino oscillations
that cannot be accommodated in the \msm.
These include the LSND signal 
\cite{LSND}, 
the MiniBooNE low-energy excess 
\cite{MiniBooNE09}, 
and neutrino-antineutrino differences 
in the MiniBooNE 
\cite{MiniBooNE2010} 
and MINOS 
\cite{MINOS2010}
experiments.
No satisfactory global description 
of these anomalies exists to date.
Here,
we focus on the possibility 
that Lorentz and CPT violation
could be responsible for 
a substantial part of the existing oscillation data,
including some or all of the anomalies.
 
Observable effects of Lorentz and CPT violation 
are conveniently described by effective field theory 
\cite{kp}.
Experimental data can be analyzed using
the Standard-Model Extension (SME)
\cite{ck},
which is the comprehensive realistic 
framework for Lorentz violation 
containing the SM and General Relativity
and incorporating CPT violation
\cite{owg}.
Numerous searches 
for nonzero coefficients for Lorentz and CPT violation
have been undertaken in recent years
using a broad range of methods 
\cite{tables}.
In the SME context,
the phenomenology of neutrino oscillations 
\cite{km,nusme1,nusme2,nusme3,nusme4,nusme5,nusme6,nusme7,%
nusme8,nusme9,nusme10,nusme11,nusme12}
and the development of techniques 
to extract limits from short- and long-baseline experiments 
\cite{kmdkm} 
have stimulated several experimental analyses 
\cite{lsndlv,MiniBooNElv,minoslv1,minoslv2,IceCube}.

In this work,
we study a simple texture $\h$ 
for the $3\times 3$ matrix governing oscillations
of three flavors of active left-handed neutrinos
that involves isotropic Lorentz and CPT violation in a chosen frame.
This requires breaking boost symmetry,
which implies neutrino mixing acquires nonstandard dependence
on the neutrino energy $E$. 
In contrast to mass terms,
which permit only a $1/E$ dependence,
isotropic Lorentz violation from effective field theory
introduces nonstandard energy dependence
even in vacuum oscillations,
a unique signal.

The texture $\h$,
which we call the `puma' model, 
was discovered by a systematic hunt 
through the jungle of possible SME-based models.
Among other criteria,
candidate textures were required to have a simple analytical form 
involving no more than three parameters.
Both mass terms and Lorentz-violating operators
of arbitrary dimension 
\cite{kmnonmin}
were included in the analysis.
The various candidates were vetted
by requiring compatibility 
with all compelling oscillation data.
Here,
we present an interesting model that 
describes established oscillation data
with only three parameters instead
of the six in the \msm. 
Remarkably,
this model also naturally reproduces
the two MiniBooNE anomalies
without extra degrees of freedom,
a feature manifestly impossible to achieve by adding to the \msm\
more neutrinos or unconventional interactions.
Moreover,
comparatively simple enhancements 
can accommodate the LSND signal and the MINOS anomaly.
The results presented here suggest 
that Lorentz- and CPT-violating models 
offer a valuable direction to pursue
in searches for simple but realistic neutrino-mixing textures.

For neutrinos,
the $3\times 3$ texture $\h$ 
can be written in the flavor basis 
and in the isotropic frame as 
\beq
\hskip -15pt
\h = A
\left(\begin{array}{ccc}
1 & 1 & 1 \\
1 & 1 & 1 \\
1 & 1 & 1 
\end{array}\right)
+B
\left(\begin{array}{ccc}
1 & 1 & 1 \\
1 & 0 & 0 \\
1 & 0 & 0 
\end{array}\right)
+C
\left(\begin{array}{ccc}
1 & 0 & 0 \\
0 & 0 & 0 \\
0 & 0 & 0 
\end{array}\right),
\quad
\label{mix}
\eeq
where 
$A(E) = {m^2}/2E$,
$B(E) = \aa\,E^2$,
and $C(E) = \cc E^5$
are real and exhibit a simple energy dependence.
Note that this texture involves
only one neutrino mass parameter $m$.
In the SME,
the coefficients $\aa$ and $\cc$
control operators for isotropic Lorentz violation
of dimension five and eight, respectively.
The operator for $\aa$ also breaks CPT,
so the $3\times3$ matrix $\hb$
governing antineutrino mixing
is obtained by reversing the sign of $\aa$
while keeping $m$ and $\cc$ unchanged. 
Since $\h$ is T invariant, 
$\aa$ also determines CP-violating effects.
For definiteness,
in this work we fix 
$m^2 = 2.6 \times 10^{-23}$ GeV$^2$,
$\aa = -2.5 \times 10^{-19}$ GeV$^{-1}$,
and $\cc = 1.0 \times 10^{-16}$ GeV$^{-4}$,
but a range of other choices 
also yields reasonable agreement with data.
The value for $m$ is consistent
with limits from direct mass measurements
and cosmological bounds
\cite{pdg}.
Intriguingly,
the texture \rf{mix} is uniformly populated at low energies,
while higher energies reveal Lorentz-violating
electron-flavor corrections
that might emerge from a unified theory at the Planck scale
\cite{ksp}.

Inspection reveals that $\h$ has a vanishing eigenvalue.
As a result,
many calculations reported here 
can be performed analytically
with comparative ease.
The eigenvalues 
$\la_1\equiv \la_-$, 
$\la_2\equiv \la_+$, 
$\la_3 \equiv 0$
of $\h$ are readily found by diagonalization 
via a unitary mixing matrix $U(E)$,
\beq
\la_\pm=\half\left[3A+B+C \pm \sqrt{(A-B-C)^2+8(A+B)^2}\right]. 
\label{eigenvals}
\eeq
The eigenvalues and mixing angles 
have nontrivial energy dependence,
a feature absent in the \msm.
The oscillation probability 
$P_{\nu_a\rightarrow\nu_b}(E)$
between flavor states $a$, $b$
has an amplitude formed from products of $U(E)$
and a phase determined by 
the dimensionless product of the baseline $L$
and the eigenvalue differences 
$\De_{a'b'}(E) = \la_{a'} - \la_{b'}$ 
between eigenstates $a'$, $b'$
\cite{km}. 
The oscillation lengths
$\le\equiv 2\pi/\De_{a'b'}(E)$
of the model \rf{mix} 
have more complicated energy dependence than those of the \msm,
which grow linearly with energy.
Since $\h$ and $\hb$ differ by the sign of $\aa$,
so do the antineutrino mixing matrix $\ueb$
\def\A5{\a{5}}
and oscillation lengths $\leb$.

\begin{figure}
\begin{center}
\centerline{\psfig{figure=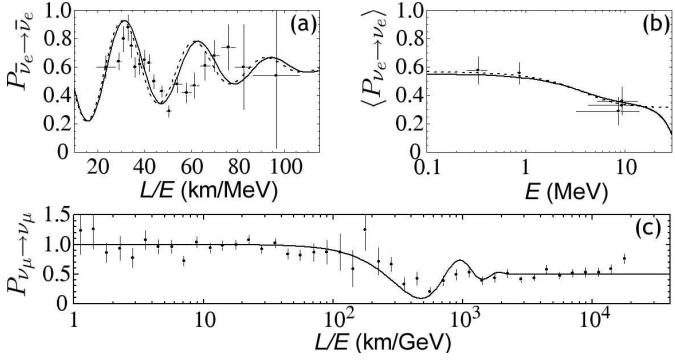,
width=\hsize}}
\caption{ \label{ModelData}
Puma model (solid) and \msm\ (dashed)
compared to (a) KamLAND \cite{KamLAND}, 
(b) solar \cite{solar}, 
and (c) SK \cite{atmospheric} data.
}
\end{center}
\end{figure}

At low energies,
the matrix \rf{mix} is dominated by the mass parameter.
As a result,
the phase of the oscillation probability 
becomes proportional to $m^2L/E$,
matching the behavior in the \msm.
The texture becomes democratic and exhibits $S_3$ invariance,
so the neutrino flavors are tribimaximally mixed. 
This guarantees consistency of the model
both with the KamLAND observation 
of long-baseline reactor-antineutrino disappearance
\cite{KamLAND}
and also with the disappearance of neutrinos
in the low-energy region of the solar-neutrino spectrum
\cite{hps}.
For solar neutrinos,
adding to $\h$ the usual effects from forward scattering in the Sun 
yields results consistent with solar data
and the \msm\ prediction
\cite{solar}.
Differences arise at energies above 30 MeV, 
which lie beyond the threshold of the solar-neutrino spectrum.
In effect,
the model eliminates the \msm\ solar mixing angle $\th_{12}$
as a degree of freedom,
with the observed mixing reproduced 
as a direct consequence of the democratic structure
of the texture \rf{mix} at low energies.

At high energies, 
the mass term $A$ in $\h$ becomes negligible.
The size and energy dependences of the terms $B$ and $C$
trigger a Lorentz-violating seesaw mechanism 
\cite{km}
that makes the eigenvalue $\la_-$
proportional to $\aa^2/\cc E$. 
The oscillation phase becomes proportional to $L/E$,
while the combination $\aa^2/\cc$ behaves like an effective mass.
The contributions from $B$ and $C$
break the low-energy $S_3$ symmetry 
to the $S_2$ subgroup in the $\mu$-$\ta$ sector.
As a result,
one eigenstate is a uniform mixture of the $\mu$ and $\ta$ flavors, 
which produces maximal mixing for atmospheric neutrinos 
in agreement with observed results 
\cite{atmospheric}.
The form of the texture \rf{mix} therefore 
also eliminates the \msm\ atmospheric mixing angle $\th_{23}$ 
as a degree of freedom.

Figure \ref{ModelData} compares 
oscillation probabilities obtained from $\h$ (solid lines)
and from the \msm\ (dashed lines)
with KamLAND, solar-neutrino, and 
Super-Kamiokande (SK) data.
The simple texture \rf{mix} 
provides a remarkable match.
This accomplishment requires 
only two of the three degrees of freedom in the model,
as described above. 
Achieving comparable results in the \msm\ 
involves four of its six degrees of freedom instead.

\begin{figure}
\begin{center}
\centerline{\psfig{figure=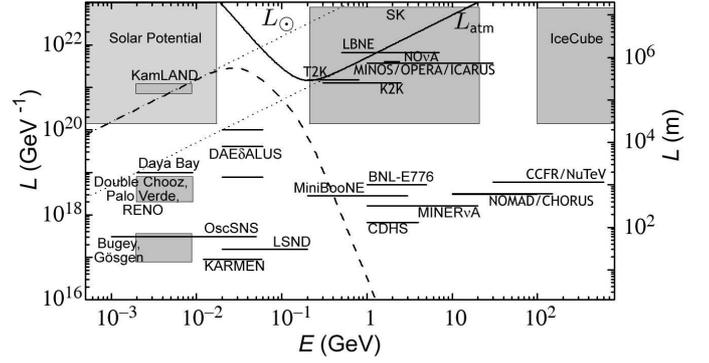,
width=\hsize}}
\caption{ \label{KMplot}
Energy dependence of oscillation lengths
in the puma model.}
\end{center}
\end{figure}

Plotting a given oscillation length 
as a function of the energy $E$
together with the relevant experimental coverage in $E$-$L$ space 
offers a powerful visual guide to oscillation signals
\cite{km}.
Provided the relevant oscillation amplitude is appreciable,
neutrino oscillations are significant 
in the region above the curve
but are mostly negligible below it.
Figure \ref{KMplot} 
displays the $E$-$L$ coverage 
of several experiments
and shows the antineutrino oscillation lengths $\leb$ 
from $\hb$ 
($\ltob$, solid curve; $\lttb$, dashed curve) 
and from the \msm\ 
($L_\odot$ and $L_{\rm atm}$, dotted straight lines).
The figure reveals that
the simple texture \rf{mix}
yields low-energy reactor oscillations 
controlled by $\lttb$,
which has the \msm\ oscillation length 
$L_\odot \propto E/\De m_\odot^2$ as an asymptote.
Also,
high-energy atmospheric oscillations
are determined by $\ltob$,
which approaches the \msm\ oscillation length 
$L_{\rm atm} \propto E/\De m_{\rm atm}^2$. 
The corresponding plot for neutrinos is similar overall,
with differences arising from the sign change of $\aa$
appearing primarily in the region 10-100 MeV.

Figure \ref{KMplot} also reveals 
that the texture \rf{mix}
is compatible with the null data 
for antineutrino disappearance 
from short-baseline reactor experiments.
The baseline $L$ for these experiments
lies well below the $\lttb$ curve,
so no oscillations occur 
even though the relevant components
of the mixing matrix $\ue$ are large.
In contrast,
the \msm\ accounts for these experimental results
as a consequence of a small mixing angle $\theta_{13}$. 
The texture \rf{mix}
is also consistent with null oscillation results 
reported by short-baseline accelerator experiments 
at high energies $\gsim 1$ GeV.
For many of these,
the mixing angles in $\ue$ essentially vanish
above 500 MeV. 
Large mixing angles in $\ue$ appear
for a subset of experiments
studying $\nu_\mu\to\nu_\ta$.
However,
no mixing is predicted 
because this oscillation channel is controlled by $\lto$,
which lies far above this region in $E$-$L$ space.

Since the texture \rf{mix} is Lorentz violating,
anisotropies must appear in any boosted frame.
Boosts of relevance to Earth-based experiments include 
the solar velocity 
$\be\simeq 10^{-3}$
relative to the cosmic microwave background,
the Earth's revolution velocity 
$\be\simeq 10^{-4}$
about the Sun,
and the tangential velocity 
$\be\simeq 10^{-5}$
of a laboratory rotating with the Earth.
These boosts generate tiny anisotropic contributions 
to $\h$ and $\hb$ 
that among other effects imply sidereal variations 
of oscillations in the laboratory frame
\cite{ak}.
In particular,
boosting $B$ introduces anisotropies 
in all the $e$ components,
while boosting $C$ introduces ones in the $ee$ component.
Of these,
only the $e\mu$ components
are experimentally constrained to date
\cite{lsndlv,MiniBooNElv,minoslv1}.
An analysis reveals that the predicted signals from $\h$ 
remain a factor of 10-100 below
the attained experimental sensitivity,
even for the maximal boost $\be\simeq 10^{-3}$.

\begin{figure}
\begin{center}
\centerline{\psfig{figure=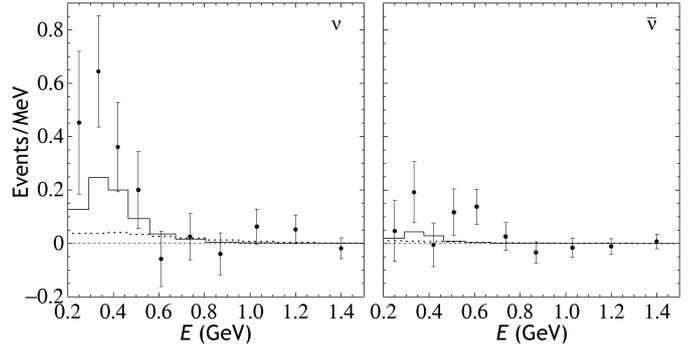,
width=\hsize}}
\caption{ \label{MiniBooNE}
Comparison of the puma model 
(solid lines; 
$\ch^2_\nu=1.0$,
$\ch^2_{\ol\nu}=0.9$),
the tandem model 
\cite{tandem} 
(dotted lines;
$\ch^2_\nu=1.9$,
$\ch^2_{\ol\nu}=1.0$),
and the \msm\ 
(dashed lines; 
$\ch^2_\nu=2.2$,
$\ch^2_{\ol\nu}=1.1$),
with MiniBooNE neutrino \cite{MiniBooNE09}
and antineutrino \cite{MiniBooNE2010} data.}
\end{center}
\end{figure}

Taken together,
the above results indicate the matrix $\h$
is compatible with confirmed experimental data
at all energies.
This match is achieved using only three real parameters,
of which two degrees of freedom 
are fixed by the existing data from accelerator experiments
and from solar- and atmospheric-neutrino measurements.
However,
the predictions of the texture \rf{mix} and of the \msm\ 
differ significantly in the range 10 MeV $\lsim E \lsim$ 1 GeV,
as can be seen in Fig.\ \ref{KMplot}. 
One pleasant surprise is that at high energies 
the small Lorentz-seesaw eigenvalue proportional to $1/E$ 
must be accompanied by a large eigenvalue
growing rapidly with $E$.
This naturally enforces a steep drop with energy 
of the length $\ltt$ relevant for $\nu_\mu\to\nu_e$ oscillations.
Our value for the third degree of freedom
permits $\ltt$ to pass through the region of sensitivity
of the MiniBooNE experiment.
Since the oscillation amplitude is large 
and decreases rapidly with energy in the same region, 
a signal is generated at energies 200-500 MeV. 
Moreover,
CPT violation involving $\aa$ makes
the oscillation signal greater for neutrinos 
than antineutrinos.
Evidence for both these features
is reported in the MiniBooNE data 
\cite{MiniBooNE09,MiniBooNE2010}.
We emphasize that these features 
emerge from the texture \rf{mix}
without additional particles, forces, or degrees of freedom.
Figure \ref{MiniBooNE} shows
results from $\h$, $\hb$, the \msm,
and the tandem model 
\cite{tandem},
which predicted a small low-energy excess prior to its discovery.
For both neutrinos and antineutrinos,
$\h$ and $\hb$ provide a better match
using a simple $\ch^2$ statistic per degree of freedom.
They also improve over the `best fits' 
obtained by varying non-\msm\ values of $\De m^2$ and $\sin^22\th$
independently for neutrinos ($\ch^2_\nu=2.6$)
\cite{MiniBooNE09}
and antineutrinos
($\ch^2_{\ol\nu}=1.1$)
\cite{MiniBooNE2010}.

Several forthcoming experiments 
using reactor antineutrinos 
and long-baseline accelerator neutrinos
are being designed to obtain precise measurements 
of the \msm\ mixing angle $\theta_{13}$
for CPT-invariant CP violation.
Among the experiments involving 
reactor-antineutrino disappearance
\cite{futurereactor},
most are insensitive to oscillations arising from $\hb$
because they are located well beneath
the $\lttb$ curve in Fig.\ \ref{KMplot}.
Analysis reveals that the largest signal from $\hb$ 
appears in the Daya Bay experiment
in the region of 2-3 MeV,
where the \msm\ predicts no oscillations. 
Other experiments plan
to investigate $\nu_\mu\to\nu_e$ 
and $\ol\nu_\mu\to\ol\nu_e$ transitions 
using long baselines of several hundred kilometers
\cite{futurelb},
for which the oscillation amplitude associated with $\h$ 
decreases rapidly above 500 MeV. 
The largest oscillation signal 
from $\h$ in this group
is about 2\% near 300 MeV in the T2K experiment,
where the \msm\ signal is below 0.5\%.
Substantial CPT violation also appears in the range 200-300 MeV,
with the $\nu_\mu$ survival probability reaching a zero minimum 
while the $\ol\nu_\mu$ survival remains above 0.8.
Another proposal is the DAE$\de$ALUS experiment
\cite{DAEdALUS},
which has a common detector 
for three $\ol\nu_\mu$ sources with different baselines.
For the currently proposed baselines,
the model \rf{mix} predicts signals 
an order of magnitude or more lower than the \msm.
However,
at a baseline of about 100 km,
$\hb$ produces a large oscillation signal 
that grows with energy,
where the \msm\ signal decreases with energy instead.

\begin{figure}
\begin{center}
\centerline{\psfig{figure=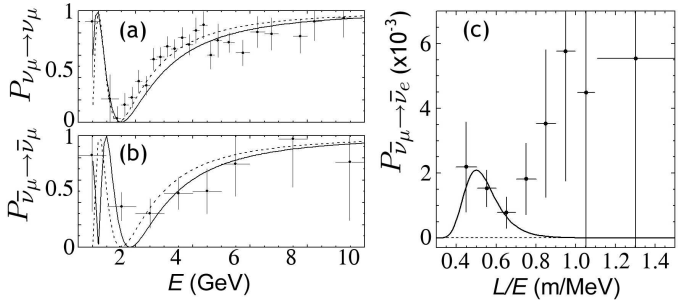,
width=\hsize}}
\caption{ \label{LM}
Comparison of texture enhancements 
of the puma model 
(solid lines; 
$\ch^2_\nu=1.4$, 
$\ch^2_{\ol\nu}=0.9$,
$\ch^2_{\rm LSND}=1.6$) 
and the \msm\ 
(dashed lines;
$\ch^2_\nu=1.0$,
$\ch^2_{\ol\nu}=1.6$,
$\ch^2_{\rm LSND}=2.6$)
with 
(a) MINOS neutrino data
\cite{atmospheric},
(b) MINOS antineutrino data 
\cite{MINOS2010},
and (c) LSND antineutrino data 
\cite{MiniBooNE09}.}
\end{center}
\end{figure}

The effectiveness of the texture \rf{mix}
in reproducing established data and satisfying constraints
suggests it offers an interesting basis 
from which to attempt the construction of a model
that also describes the antineutrino-oscillation anomalies 
reported by the LSND
\cite{LSND} 
and MINOS
\cite{MINOS2010}
experiments.
One simple texture-preserving extension 
of the three-parameter model 
involves an additional contribution  
to the $e$-$\mu$ and $e$-$\tau$ sectors of the form $\cc'E$,
where $\cc'$ is a coefficient for CPT-even Lorentz violation 
chosen here for definiteness as $\cc'=2.0\times10^{-20}$. 
The presence of this fourth degree of freedom 
leaves unaffected the main features of the three-parameter model
but combines with the coefficient $\aa$
to trigger differing oscillation probabilities
for neutrinos and antineutrinos at high energies. 
The difference appears as a relative shift between 
the minima of the survival probabilities
for neutrinos and antineutrinos.
This is consistent with the MINOS anomaly,
as can be seen in Fig.\ \ref{LM}. 
Similar shifts are also predicted
in future long-baseline accelerator experiments
\cite{futurelb},
with a notable peak in the T2K antineutrino data
where the \msm\ predicts a minimum.
We emphasize that these CPT-violating effects are achieved 
without neutrino-antineutrino mass differences
and are consistent with effective field theory
\cite{owg}.
Indeed,
the only mass parameter in the model
appears in the original texture $\h$,
and it is negligible at MINOS energies.

The LSND signal \cite{LSND}
is absent from the three-parameter model $\hb$
because the baseline lies far below 
the $\lttb$ curve in Fig.\ \ref{KMplot}.
However,
another option for enhancing the model 
is an energy-localized modification 
that preserves compatibility of the texture with established data. 
We consider here a simple 
three-parameter gaussian enhancement 
$\de h^{\ol\nu}_{\rm eff} = \al \exp{[-\be(E - \ep)^2]}$
for the $\ol e$-$\ol \mu$ and $\ol e$-$\ol \ta$ sectors of $\hb$.
The CPT-conjugate enhancement is obtained by 
changing the signs of $\al$ and $\ep$,
which maps its effects outside the physical range
and so leaves neutrino oscillations unaffected.
Introducing an enhancement 
$\de h^{\ol\nu}_{\rm eff}$
can produce a localized valley in the $\lttb$ curve
with minimum approaching the region of LSND sensitivity. 
For example, 
Fig.\ \ref{LM} shows the signal
from a $\de h^{\ol\nu}_{\rm eff}$
centered at $\ep =60$ MeV 
with amplitude $\al = 3.0\times 10^{-19}$ GeV
and width $\be = 3.0\times 10^{3}$ GeV$^{-2}$.
No effects arise in other existing experiments
because the valley is localized in $E$,
while sidereal variations lie below current limits.
However,
this enhancement predicts
a nonzero probability for the planned OscSNS experiment
\cite{OscSNS},
growing to about 0.5\% at 50 MeV. 
It also predicts large effects 
from all three sources in the proposed DAE$\de$ALUS experiment
\cite{DAEdALUS},
with a striking signal from the near source
that is about 100 times larger than the \msm\ expectation.

\begin{figure}
\begin{center}
\centerline{\psfig{figure=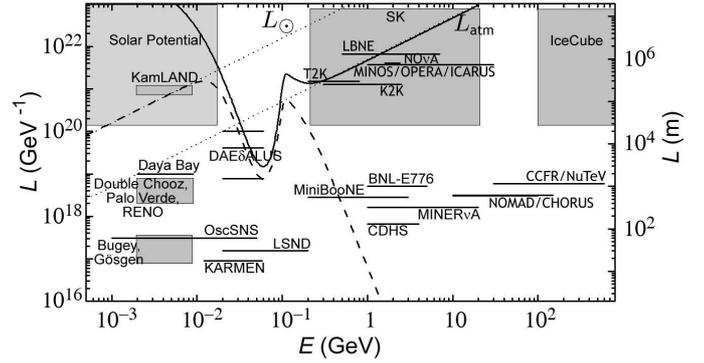,
width=\hsize}}
\caption{ \label{KMplot2}
Energy dependence of oscillation lengths
in the doubly enhanced puma model.}
\end{center}
\end{figure}

The above extension and enhancement lie in disjoint energy regions,
so we can incorporate both simultaneously.
The result is a seven-parameter doubly enhanced texture
that appears globally compatible with all compelling data
and the various existing anomalies. 
The changes to the eigenvalues \rf{eigenvals} of $\h$
arising from the double enhancement
are found by the substitutions
$B\to B +\de h$, $C\to C - \de h$
with $\de h = \cc'E - \al \exp{[-\be(E + \ep)^2]}$.
The effect on the oscillation lengths
is shown in Fig.\ \ref{KMplot2}.
Since the anomalies remain to be confirmed experimentally
and since the doubly enhanced texture 
requires four more parameters than the frugal three of $\h$,
we interpret this construction primarily as an existence proof 
revealing the surprising effectiveness
of simple Lorentz-violating textures.
However,
any putative enhancement of the \msm\ 
will also require extra degrees of freedom,  
and the seven-parameter texture does appear 
at present to offer the only global description 
incorporating all anomalies.
Further tests of these ideas
can be expected in the near future
from additional experimental data 
elucidating the various anomalies.

This work was supported in part
by the Department of Energy
under grant DE-FG02-91ER40661
and by the Indiana University Center for Spacetime Symmetries.


\begin{thebibliography}{99}

\bibitem{pdg}
K.\ Nakamura \etal,
J.\ Phys.\ G {\bf 37}, 075021 (2010).

\bibitem{LSND}   
A.\ Aguilar \etal,
Phys.\ Rev.\ D {\bf 64}, 112007 (2001).
  
\bibitem{MiniBooNE09}   
A.A.\ Aguilar-Arevalo \etal, 
Phys.\ Rev.\ Lett.\ {\bf 102}, 101802 (2009).

\bibitem{MiniBooNE2010}
A.A.\ Aguilar-Arevalo \etal, 
Phys.\ Rev.\ Lett.\ {\bf 105}, 181801 (2010).

\bibitem{MINOS2010}   
P.\ Vahle, 
Nucl.\ Phys.\ B Proc.\ Suppl., in press. 

\bibitem{kp}
V.A.\ Kosteleck\'y and R.\ Potting,
Phys.\ Rev.\ D {\bf 51}, 3923 (1995).

\bibitem{ck}
D.\ Colladay and V.A.\ Kosteleck\'y,
Phys.\ Rev.\ D {\bf 55}, 6760 (1997);
Phys.\ Rev.\ D {\bf 58}, 116002 (1998);
V.A.\ Kosteleck\'y,
Phys.\ Rev.\ D {\bf 69}, 105009 (2004).

\bibitem{owg}
O.W.\ Greenberg,
Phys.\ Rev.\ Lett.\ {\bf 89}, 231602 (2002).

\bibitem{tables}
V.A.\ Kosteleck\'y and N.\ Russell,
Rev.\ Mod.\ Phys.\ {\bf 83}, 11 (2011).

\bibitem{km}
V.A.\ Kosteleck\'y and M.\ Mewes,
Phys.\ Rev.\ D {\bf 69}, 016005 (2004);
Phys.\ Rev.\ D {\bf 70}, 031902 (R) (2004).

\bibitem{nusme1}
V.\ Barger, D.\ Marfatia, and K.\ Whisnant,
Phys.\ Lett.\ B {\bf 653}, 267 (2007).

\bibitem{nusme2}
N.\ Cipriano Ribeiro \etal,
Phys.\ Rev.\ D {\bf 77}, 073007 (2008).

\bibitem{nusme3}
A.E.\ Bernardini and O.\ Bertolami,
Phys.\ Rev.\ D {\bf 77}, 085032 (2008).

\bibitem{nusme4}
B.\ Altschul,
J.\ Phys.\ Conf.\ Ser. {\bf 173} 012003 (2009).

\bibitem{nusme5}
S.\ Hollenberg, O.\ Micu, and H.\ P\"as,
Phys.\ Rev.\ D {\bf 80}, 053010 (2009).

\bibitem{nusme6}
S.\ Ando, M.\ Kamionkowski, and I.\ Mocioiu,
Phys.\ Rev.\ D {\bf 80}, 123522 (2009).

\bibitem{nusme7}
M.\ Bustamante, A.M.\ Gago, and C.\ Pe\~na-Garay,
J.\ Phys.\ Conf.\ Ser.\ {\bf 171}, 012048 (2009).

\bibitem{nusme8}
P.\ Arias and J.\ Gamboa,
Int.\ J.\ Mod.\ Phys.\ A {\bf 25}, 277 (2010).

\bibitem{nusme9}
S.\ Yang and B.-Q.\ Ma,
Int.\ J.\ Mod.\ Phys.\ A {\bf 24}, 5861 (2009).

\bibitem{nusme10}
D.M.\ Mattingly \etal,
JCAP {\bf 1002}, 007 (2010).

\bibitem{nusme11}
A.\ Bhattacharya \etal,
JCAP {\bf 1009}, 009 (2010).

\bibitem{nusme12}
C.M.\ Ho, arXiv:1012.1053.

\bibitem{kmdkm}
J.S.\ D\'iaz \etal,
Phys.\ Rev.\ D {\bf 80}, 076007 (2009);
V.A.\ Kosteleck\'y and M.\ Mewes,
Phys.\ Rev.\ D {\bf 70}, 076002 (2004).

\bibitem{lsndlv}
L.B.\ Auerbach \etal,
Phys.\ Rev.\ D {\bf 72}, 076004 (2005).

\bibitem{MiniBooNElv}
T.\ Katori,
arXiv:1008.0906.

\bibitem{minoslv1}
P.\ Adamson \etal,
Phys.\ Rev.\ Lett.\ {\bf 101}, 151601 (2008).

\bibitem{minoslv2}
P.\ Adamson \etal,
Phys.\ Rev.\ Lett.\ {\bf 105}, 151601 (2010).

\bibitem{IceCube}
R.\ Abbasi \etal,
Phys.\ Rev.\ D {\bf 82}, 112003 (2010).

\bibitem{kmnonmin}
V.A.\ Kosteleck\'y and M.\ Mewes,
Phys.\ Rev.\ D {\bf 80}, 015020 (2009);
in preparation.

\bibitem{ksp}
V.A.\ Kosteleck\'y and S.\ Samuel,
Phys.\ Rev.\ D {\bf 39}, 683 (1989);
V.A.\ Kosteleck\'y and R.\ Potting,
Nucl.\ Phys.\ B {\bf 359}, 545 (1991).

\bibitem{KamLAND}   
S.\ Abe {\it et al.}, 
Phys.\ Rev.\ Lett.\ {\bf 100}, 221803 (2008).

\bibitem{hps}
P.F.\ Harrison, D.H.\ Perkins, and W.G.\ Scott,
Phys.\ Lett.\ B {\bf 530}, 167 (2002).

\bibitem{solar}   
G.\ Bellini \etal, 
Phys.\ Rev.\ D {\bf 82}, 033006 (2010).

\bibitem{atmospheric}   
M.H.\ Ahn \etal,
Phys.\ Rev.\ Lett.\ {\bf 90}, 041801 (2003);
Y.\ Ashie \etal,
Phys.\ Rev.\ Lett.\ {\bf 93}, 101801 (2004);
D.G.\ Michael \etal,
Phys.\ Rev.\ Lett.\ {\bf 97}, 191801 (2006).

\bibitem{ak}
V.A.\ Kosteleck\'y,
Phys.\ Rev.\ Lett.\ {\bf 80}, 1818 (1998).

\bibitem{tandem}   
T.\ Katori \etal,
Phys.\ Rev.\ D {\bf 74}, 105009 (2006). 

\bibitem{futurereactor}
J.K.\ Ahn \etal,
arXiv:1003.1391;
X.\ Guo \etal,
hep-ex/0701029;
F.\ Ardellier \etal,
hep-ex/0606025.

\bibitem{futurelb}
Y.\ Itow \etal,
hep-ex/0106019;
V.\ Barger \etal,
arXiv:0705.4396;
D.S.\ Ayres \etal,
hep-ex/0503053.

\bibitem{DAEdALUS}     
J.\ Alonso \etal,
arXiv:1006.0260.

\bibitem{OscSNS}
G.T.\ Garvey \etal,
Phys.\ Rev.\ D {\bf 72}, 092001 (2005).

\end{thebibliography}
\end{document}